# Unconventional band splitting of CeSb in the devil's staircase transition

Tongrui Li[1*], Zhanfeng Liu[1*], Peng Li[2*,†], Yuzhe Wang[1, 3], Zhisheng Zhao[1, 3], Shiwu Su[1, 3], Zhicheng Jiang[1], Yuhao Hong[1], Hui Tian[1], Xin Zheng[1], Yi Liu[1], Yilin Wang[1, 3], Zhengtai Liu[4], Dawei Shen[1, 5], Zhe Sun[1†], Yang Liu[6†], Juan Jiang[1, 3†], Donglai Feng[1, 5]

[1] *National Synchrotron Radiation Laboratory and School of Emerging Technology, University of Science and Technology of China, Hefei 230026, People's Republic of China*
[2] *Quantum Science Center of Guangdong–Hong Kong–Macao Greater Bay Area (Guangdong), Shenzhen 518045, China*
[3] *Hefei National Laboratory, University of Science and Technology of China, Hefei 230088, China*
[4] *Shanghai Synchrotron Radiation Facility, Shanghai Advanced Research Institute, Chinese Academy of Sciences, Shanghai, 201210, China*
[5] *School of Nuclear Science and Technology, and New Cornerstone Science Laboratory, University of Science and Technology of China, Hefei, 230026, China*
[6] *Center for Correlated Matter and School of Physics, Zhejiang University, Hangzhou 310058, China*

*These authors contributed equally to this work.

[†]Emails: lipeng@quantumsc.cn; zsun@ustc.edu.cn; yangliuphys@zju.edu.cn; jjiangcindy@ustc.edu.cn

**Abstract**

The interplay between magnetism and electronic band structure is a central theme in condensed matter physics. CeSb, with its complex devil's staircase antiferromagnetic transition, offers a unique opportunity to explore this interplay. Using angle-resolved photoemission spectroscopy (ARPES), we investigate the electronic structure evolution across the devil's staircase transition. Upon entering the antiferromagnetic phase, we observe an intriguing band splitting of the electron pocket around the X point. The energy separation between the split bands changes abruptly with temperature, consistent with the characteristics of the first-order phase transition. However, their respective spectral weights behave gradually with temperature. Combined with our density functional theory (DFT) calculations, we suggest that this atypical behavior deviates from conventional magnetically induced band splitting and potentially arises from the intricate modulation of paramagnetic and antiferromagnetic layers within the devil's staircase transition. Our results provide insights into the complex relationship between electronic structure and magnetism in correlated electron systems.

**Introduction**

In strongly correlated systems, the interplay of multiple degrees of freedom, such as electronic charge, spin, and orbital, gives rise to complex and unexpected phenomena. Of particular interest are 4$f$ electron systems, which exhibit a range of captivating phenomena, including heavy fermion superconductivity [1-3], the Kondo effect [4-7], and quantum criticality [7]. Understanding the magnetic mechanisms in such systems and unraveling their connections to electronic structures have become a focus of scientific investigation. The RePn (Re = La, Ce, Pr, Nd, Sm, Gd, Pn = Sb, Bi) family of compounds exhibit a wide range of interesting properties, including large magnetoresistance [8-11] and topological properties in the paramagnetic (PM) phases [12-16]. Recently, the antiferromagnetic (AFM) phases of these compounds have attracted considerable attention [17-21] due to the intimate connection between magnetism and electronic structure. For instance, NdBi [17-19] and NdSb [20-22] exhibit intriguing Fermi-arc surface states and unconventional magnetic splitting in their antiferromagnetic phases related to the multi-q magnetic order. CeBi demonstrates unexpected band structure changes in the antiferromagnetic state[23].

CeSb stands out due to its exceptionally complex magnetic phase diagram, featuring more than seven distinct antiferroparamagnetic (AFP) phase transitions below its Néel temperature ($T_N$) of 17 K. This intricate sequence of transitions, known as the devil's staircase [24-26], includes a type-IA (↑↑↓↓) AFM structure below 8 K [27]. This structure consists of ferromagnetic aligned planes of Ce moments stacked antiferromagnetically along the c-axis (↑↑↓↓), where the arrows represent the magnetization direction within each plane. Previous ARPES studies on CeSb within the 'devil's staircase' regime have reported significant band renormalization of the hole bands near the Brillouin zone (BZ) center [28,29], initially attributed to band folding. Another ARPES study observed splitting-like electron bands at the BZ corner in the AFM phases [30], which was interpreted as Zeeman-type splitting caused by in-plane ferromagnetic exchange within a two-channel model. While neutron scattering studies have provided insights into the evolution of nonmagnetic Ce layers within the devil's staircase [24,26,31], direct electronic evidence supporting this model remain elusive.

In this letter, we use ARPES measurements and first-principles calculations to investigate the electronic band structure evolution of CeSb across the devil's staircase transition. We identify two distinct magnetic domains, differentiated by the electronic structure around the Γ point, and both present an unconventional band splitting at the X point. While the energy separation between the split bands remains largely temperature independent, their spectral weights show a pronounced temperature dependence. Combined with our DFT calculations, we propose that this splitting in the electron bands arises from the superposition of bands originating from both the nonmagnetic and magnetic Ce layers within the modulated devil's staircase magnetic structure.

**Results and Discussion**

The paramagnetic phase of CeSb crystallizes in the NaCl structure (#Space Group F$\bar{m}$3m). The magnetic structure of the type-IA AFM is shown in Fig. 1a, with magnetic moments aligned along the c-axis. Experimentally, there are two distinct magnetic domains based on the two different cleavage planes: the *c*-domain, with out-of-plane moments and the *ab*-domain, with in-plane moments. Figure 1b illustrates the three-dimensional and surface-projected BZs of both the PM and AFM phases. Figure 1c shows the ARPES spectra along the high symmetry direction $\bar{\Gamma}\bar{X}_{PM}$ in the PM state, which agrees well with the calculated band structure in Fig. 1d and previous reports [32,33] (see Supplementary Figure 1 for more details of the Fermi surface). The energy distribution curve (EDC) at the $\bar{X}_{PM}$ point presents a clear gap feature, indicating a topologically trivial PM phase [33]. While the overall Fermi surface topologies of the two antiferromagnetic domains appear similar (Figs. 1e and 1f), distinct differences emerge in the detailed band dispersions. Figures 1g and 1h show ARPES spectra along $\bar{\Gamma}\bar{X}_{PM}$ for the *c*-domain and *ab*-domain, respectively (cut directions indicated by black solid lines in Figs. 1e and 1f). The *c*-domain exhibits complex reconstructed bands near the $\bar{\Gamma}$ point, while the *ab*-domain shows fewer bands without clear electron-like features. These distinct characteristics are consistent with the previous report [28] (see Supplementary Figure 2 and Supplementary Figure 3 for further details).

To further investigate the electronic structure of the AFM phase, we conducted low-temperature ARPES measurements on both magnetic domains. Figure 2a presents the FS map of the *ab*-domain in the $k_x$-$k_z$ plane measured using 28 eV photons (see Supplementary Figure 4 for the determination of different magnetic domains). The relationship between photon energy and $k_z$ is shown in Fig. 2b. Figures 2c and 2d display the ARPES spectra and the corresponding second derivative images along cut 1 as indicated in Fig. 2a at 6 K (T<$T_N$) and 19.2 K (T>$T_N$), respectively. The parabolic electron band clearly splits in the AFM state, while the V-shaped electron band remains unaffected. On the contrary, along cut 2 (Figures 2e-2g), the V-shaped band exhibits splitting, whereas the parabolic band does not. Notably, a crossing-like feature shows up in Fig. 2e, likely arising from bulk bands as it disappears at a different photon energy (Fig. 2f). These observations indicate that the band splitting in the *ab*-domain exhibits $C_2$ symmetry. It is important to note that, although RePn system exhibits a strong *kz* broadening effect [11,12,20], making it difficult to detect a pure *kz* even with a single photon energy, the observed electronic structure that varies with photon energy reflects the intrinsic characteristics of the bulk bands. Figure 2h shows the FS of the *c*-domain, and Figs. 2i-k present ARPES spectra and second derivative images along cut 3 (indicated in Fig. 2h) at different photon energies and temperatures. Unlike the *ab*-domain, both the V-shaped and parabolic bands exhibit splitting in the *c*-domain, consistent with the $C_4$ symmetry. The weaker splitting of the V-shaped bands in the *c*-domain may be partially obscured by $k_z$-broadening arising from the finite escape depth of the photoelectrons (see Supplementary Figure 4 for the complete dataset).

To understand the origin of the band splitting in ARPES spectra, we performed DFT calculations for the AFM phase. The Hubbard U parameter was determined based on the energy position of the Ce $4f^0$ level. Figure 3a shows the resonant photoemission measurement at the Ce $4d \rightarrow 4f$ edge (122 eV) and reveals the $4f^0$ level at a binding energy between 2.6 eV and 3.2 eV. With *f* electrons treated as valence electrons, U = 4.5 eV and J = 0.7 eV were determined (see Supplementary Figure 5 for details). Figure 3b plots the band calculations for the AFM ground state (type-IA), with the energy position of the $4f^0$ level accurately reproduced. However, this calculation (Fig. 3d) does not reproduce the observed band splitting at the $\overline{M}(\overline{X}_{PM})$ point (Fig. 3c), where the experimentally measured splitting is approximately 0.11 eV. We also explored other antiferromagnetic configurations, but none of them can reproduce the ARPES measurements (see Supplementary Figure 6 for more magnetic configurations). Calculations for a ferromagnetic configuration (Fig. 3e) yield a Zeeman splitting of approximately 0.075 eV at the $\overline{M}(\overline{X}_{PM})$ point, smaller than the experimental value. Furthermore, the calculated band gap between the top of the valence bands and the bottom of the conduction bands at the $\overline{M}(\overline{X}_{PM})$ point in the ferromagnetic phase is ~ 0.17 eV, significantly larger than the experimental value. Thus, the possible ferromagnetic bands cannot reconcile with ARPES data. Interestingly, superimposing the calculated band structures of the paramagnetic and antiferromagnetic states (Fig. 3f) yields a splitting of approximately 0.12 eV between the upper and lower parabolic electron bands at the $\overline{M}(\overline{X}_{PM})$ point, in close agreement with the experimental value. Figures 3g, h, and i show the superposition of the calculated results (from Figures 3d, e, and f) and the experimental results (from Figures 3c) respectively. It is evident that the calculations in Figure 3i align well with the experimental data. This suggests that the observed splitting may arise from a superposition of paramagnetic and antiferromagnetic contributions. We summarize our findings in the schematic diagram shown in Supplementary Figure 7. It is worth noting that we employ the superposition method to provide a proposal, as this approach better aligns with the experimental results. However, this superposition approach may not fully capture the underlying physics. The devil's staircase implies a complex spatial modulation of magnetic order, with potentially sharp interfaces between paramagnetic and antiferromagnetic regions, thus the temperature evolution of the band splitting would give more spectroscopic evidences.

Due to the use of vacuum ultraviolet light, the electron escape depth is limited and cannot effectively reach a significant depth. To better investigate the temperature evolution of the electronic structure of the devil's staircase transition, we will focus on exploring the temperature dependent experiment in the *ab* domain. Figure 4a shows ARPES spectra along cut 2 (Fig. 2a) for sample #1, and Fig. 4c shows similar data for sample #2 measured with a He lamp (21.2 eV). Both datasets reveal a splitting of the V-shaped electron band in the AFM phases, similar to previous reports [30]. The inner and outer bands are indicated by arrows in Figs. 4a and 4c, and are also evident in the momentum distribution curve (MDC) at $E_F$. Figs. 4b and 4d show EDCs extracted at the $\overline{X}_{PM}$ point, fitted with two peaks (see Supplementary Figure 8 for fitting details). Below 17 K, two distinct peaks are observed, corresponding to the inner (blue dots) and outer (red dots) branches of the split band. As the temperature decreases, the peak position gradually shifts toward lower energy levels, resembling the phenomenon observed in CeBi [35], this behavior may be associated with the hybridization of *f* electrons and conduction electrons near the Fermi level. In addition, the spectral weight of the inner band gradually increases, while that of the outer band gradually diminishes. This temperature dependence correlates with the evolution of the devil's staircase, as shown in Fig. 4e, where the fraction of nonmagnetic Ce layers (gray dots) decreases with decreasing temperature, as determined by neutron scattering [24,26,31]. This suggests a possible connection between the observed band splitting and the presence of both magnetic and nonmagnetic Ce layers within the devil's staircase structure. Additionally, we attempted to calculate the electronic structure in intermediate magnetic configurations, such as the AFP1 state; however, we were unable to obtain the non-magnetic layer of Ce in the DFT calculations.

We quantified the temperature dependence of the outer band intensity by calculating the ratio $I_{outer} = outer/(inner + outer)$, where *inner* and *outer* are the intensities of the inner and outer peaks in the EDCs, respectively. Figure 4f shows $I_{outer}$ for three different samples (#1, #2, and #3; see Supplementary Figure 9 for additional data of #3) alongside the temperature-dependent fraction of nonmagnetic Ce layers, $I_{PM}$, within the devil's staircase structure. The close agreement between $I_{outer}$ and the $I_{PM}$ suggests that the outer band originate from the nonmagnetic Ce layers. Remarkably, the energy separation between the inner and outer bands remains constant at approximately 0.1 eV below 17 K, as shown in Fig. 4g. These observations contrast with the behavior of typical magnetically induced band splitting, such as that observed in EuB$_6$ [36], where the splitting increases with decreasing temperature. However, this difference does not definitively rule out a magnetic origin for the splitting in CeSb, as the underlying mechanisms may be different. The observed temperature-dependent spectral weight transfer between the inner and outer bands is consistent with the changing fraction of magnetic and nonmagnetic Ce layers within the devil's staircase structure [37,38]. The residual intensity of the outer band below 8 K, where the devil's staircase model predicts only magnetic Ce layers, likely arises from phase fluctuations of the magnetic structure. This is because the lowest experimentally measured temperature is quite close to the phase transition temperature. Our data strongly suggest that the apparent band splitting in CeSb arises from the coexistence of paramagnetic and antiferromagnetic states, providing direct spectroscopic evidence for the devil's staircase picture.

**Conclusions**

In summary, our combined ARPES and DFT + U study provides insights into the electronic structure of CeSb across the devil's staircase transition. The apparent band splitting observed at the $\bar{X}_{PM}$ point in antiferromagnetic phase arises from the coexistence of paramagnetic and AFM states, reflecting the complex magnetic structure of CeSb. The temperature-dependent spectra weight transfer between the split bands can be well explained by the changing fraction of magnetic and nonmagnetic Ce layers within the devil's staircase magnetic structures. Our results provide direct spectroscopic evidence of the existence and evolution of the nonmagnetic Ce layers, offering a deeper understanding of the unusual electronic properties of this material.

**Methods**

Single crystals of CeSb were synthesized using the indium flux method with a molar ratio of Ce: Sb: Sn of 1: 1: 20. The starting materials were weighted and loaded in alumina crucibles, which were then sealed in an evacuated quartz tube and heated to 1150 ℃ before being cooled down to 800 ℃. Finally, the samples were separated from the indium in a centrifuge. The typical crystal size was 4 × 4 × 4 mm.

ARPES measurements were performed at beamline 03U of the Shanghai Synchrotron Radiation Facility (SSRF) in China. The ultra-high vacuum measurement was maintained below 8 × $10^{-11}$ Torr, and data were recorded using a Scienta DA30 analyzer. The total convolved energy and angle resolutions were 10 meV and 0.2°, respectively. The fresh surface for ARPES measurements was obtained by cleaving the CeSb samples in-situ along its natural cleavage plane.

Bulk electronic structure calculations were performed using density functional theory (DFT) with a plane wave basis projected augmented wave method [39], as implemented in the Vienna *ab-initio* simulation package (VASP) [40]. To address the overestimation of bandwidth by the PBE functional, we slightly increased the lattice constant by a factor of 1.08, following the method outlined in Ref. [20]. This adjustment ensured that no band inversion occurs at the X point in the paramagnetic state. The generalized gradient approximation of Perdew-Burke-Ernzerhof (PBE) [41] was used as the exchange-correlation potential. A tight binding model based on Wannier functions [42,43] was constructed to reproduce the spectral functions with the selection of Ce *d* and Sb *p* orbitals in the paramagnetic state. The spectrum and Fermi surfaces were calculated using surface Green's function methods [44,45], as implemented in WannierTools [46]. The 4f electrons were treated as valence electrons and spin-orbit coupling was included in the antiferromagnetic state. The energy cutoff was set to 300 eV and a 9 × 9 × 7 Γ-centered k-mesh was employed in the calculation. U = 4.5 eV and J = 0.7 eV were used for our DFT+U+SOC calculations in the AFM phase, following the method of Liechtenstein *et al* [47], which matched well with the energy position of the $4f^0$ in resonance ARPES measurements. Fermi levels were shifted up by 180 meV to align with the ARPES spectra.

**Data availability**

The data that support the findings of this study are available from the corresponding authors upon request.

## Acknowledgments

This work is supported by the National Key R&D Program of China (Grant No. 2023YFA1406304 (J. J), No. 2022YFA1402200 (Y. L)), the National Natural Science Foundation of China (Grant No. 12174362 (J. J), No. 11790312 (D. L. F), No. 92065202 (J. J)), the Postdoctoral Fellowship Program of CPSF(GZC20232530) (Z. F. L), the Innovation Program for Quantum Science and Technology (No. 2021ZD0302803 (D. L. F)) and the New Cornerstone Science Foundation (D. L. F). Part of this research used Beamline 03U of the Shanghai Synchrotron Radiation Facility, which is supported by ME2 project under contract no. 11227902 from National Natural Science Foundation of China.

## Author Contributions

T.R.L., P.L. and Z.F.L. contribute equally to this work. P.L. and J.J. conceived the experiments. T.R.L., P.L. and J.J. carried out ARPES measurements with the assistance of Z.C.J., Yi. L., Z.S., Z.T.L. and D.W.S., T.R.L. performed the DFT calculations under the supervision of Y.L.W., Z.F.L. and P.L. synthesized and characterized bulk single crystals. Y.Z.W., Z.S.Z., S.W.S., Y.H.H., H.T., X.Z. contributed to measurements and data analysis. T.R.L., P.L., Z.S., Yang. L., J.J. and D.L.F. wrote the manuscript. All authors contributed to the scientific planning and discussions.

## Competing interests

The authors declare no competing interests. Dawei Shen is an Editorial Board Member for Communications Materials and was not involved in the editorial review of, or the decision to publish, this Article.

## Additional information

**Supplementary information** The online version contains supplementary material available at xxxx.

**Correspondence** and requests for materials should be addressed to Juan Jiang or Peng Li.

**Reprints and permissions information** is available at http://www.nature.com/reprints

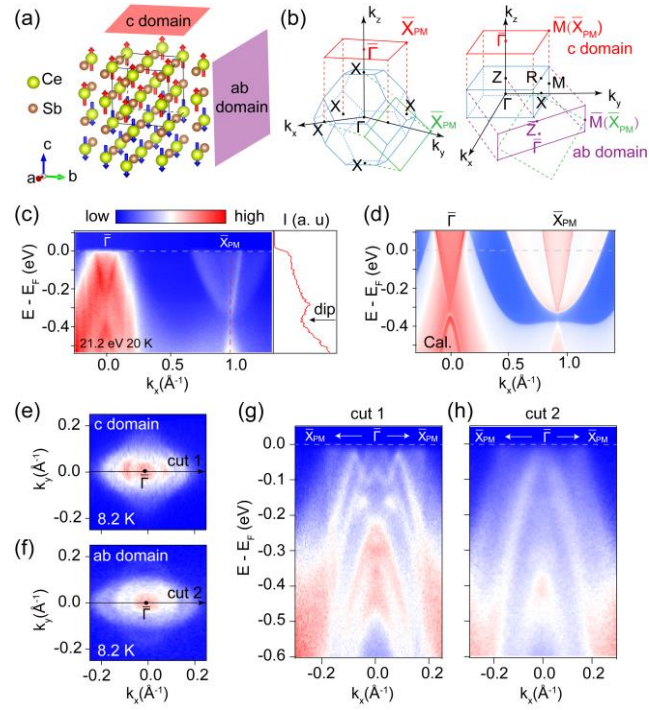

**Fig. 1 Electronic structure of paramagnetic state and two magnetic domains of the antiferromagnetic ground state near $\bar{\Gamma}$.** **(a)** Arrangement of magnetic moments of the type-IA antiferromagnetic structure, illustrating two possible magnetic domains: the *c*-domain and the *ab*-domain. **(b)** Brillouin zones of the paramagnetic phase and the type-IA antiferromagnetic phase. **(c)** ARPES intensity along $\bar{\Gamma}\bar{X}_{PM}$ in the paramagnetic phase. The EDC obtained along the red dashed line is plotted on the right side, revealing a distinct band gap. **(d)** Calculated band structure along $\bar{\Gamma}\bar{X}_{PM}$ in the paramagnetic phase. **(e)** The Fermi surface map of the *c*-domain near the $\bar{\Gamma}$ point. **(f)** The Fermi surface map of the *ab*-domain near the $\bar{\Gamma}$ point. **(g, h)** ARPES intensity plots along $\bar{\Gamma}\bar{X}_{PM}$ of cut 1 in **(e)** and cut 2 in **(f)**, respectively.

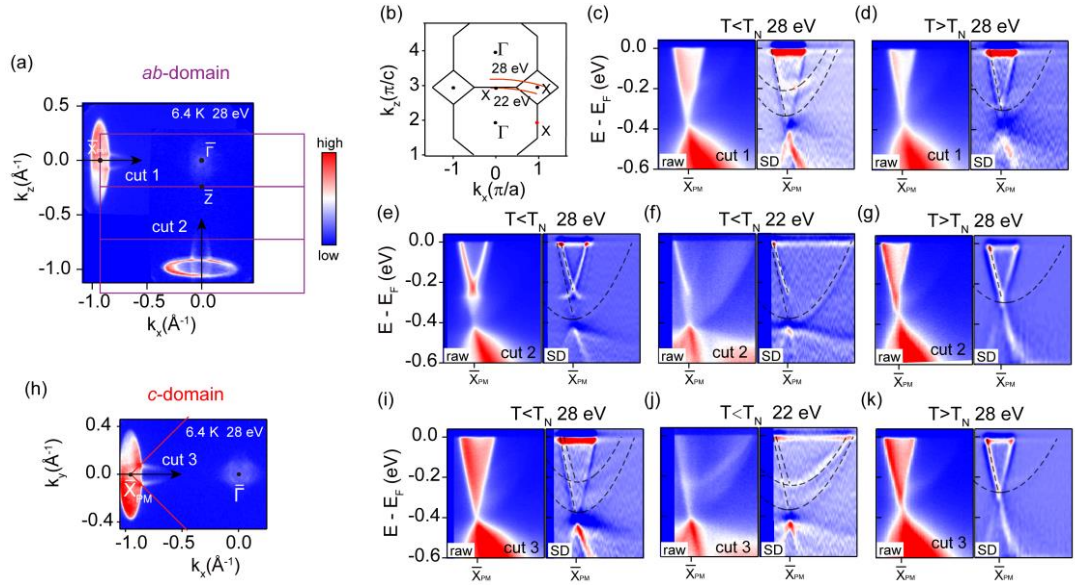

**Fig. 2 Observation of splitting-like electron bands in *ab*- and *c*-domains. (a)** Fermi surface map of the *ab*-domain measured using 28 eV photons at T = 6.4 K. **(b)** Relationship between photon energy and $k_z$ in the $k_z$-$k_x$ plane. **(c, d)** ARPES spectra along $\bar{\bar{\Gamma}}\bar{\bar{X}}_{PM}$ of cut 1 in **(a)** and the corresponding second derivative images, measured using 28 eV photons at T= 6 K and 19.2 K, respectively. **(e, f, g)** ARPES spectra along $\bar{\bar{\Gamma}}\bar{\bar{X}}_{PM}$ of cut 2 in **(a)** and the corresponding second derivative spectra, measured at 6.4 K with 28 eV and 22 eV photons, and 19.2 K with 28 eV photons, respectively. **(h)** Fermi surface map of the *c*-domain at 6.4 K with 28 eV photons. **(i, j, k)** ARPES spectra along $\bar{\bar{\Gamma}}\bar{\bar{X}}_{PM}$ of cut 3 in **(h)** and the corresponding second derivative spectra, measured at 8.1 K with 28 eV photons, 6.6 K with 22 eV photons, and 19.2 K with 28 eV photons, respectively.

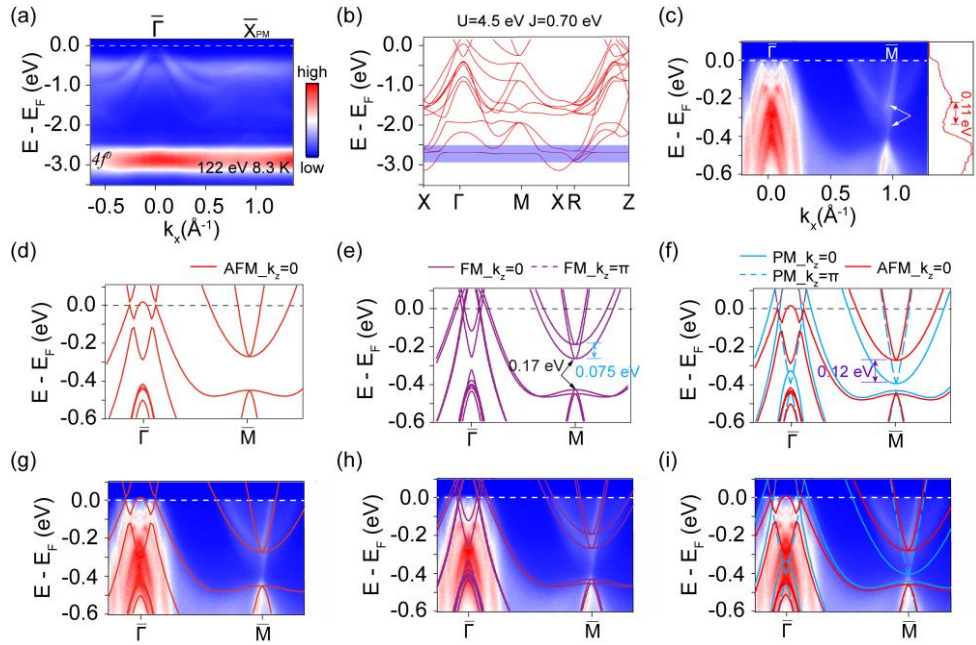

**Fig. 3 Identification of paramagnetic and antiferromagnetic layers using DFT + U calculations.**
**(a)** Ce $4f^0$ energy level determined from resonant ARPES measurements using 122 eV photons, located around 2.8 eV below the Fermi level. **(b)** Values of the Hubbard U and Hund's coupling J parameters for the DFT+U calculations of the antiferromagnetic ground phase. **(c)** ARPES spectra of the *c*-domain at 8.2 K. **(d)** Calculated band structure for the type-IA AFM phase. **(e)** Calculated band structure for the ferromagnetic phase. **(f)** Superimposed DFT calculations of the type-IA AFM and PM phases. The color scale represents the intensity of ARPES spectra. **(g, h, i)** ARPES spectra overlapped with DFT calculations in **(d, e, f)**, respectively.

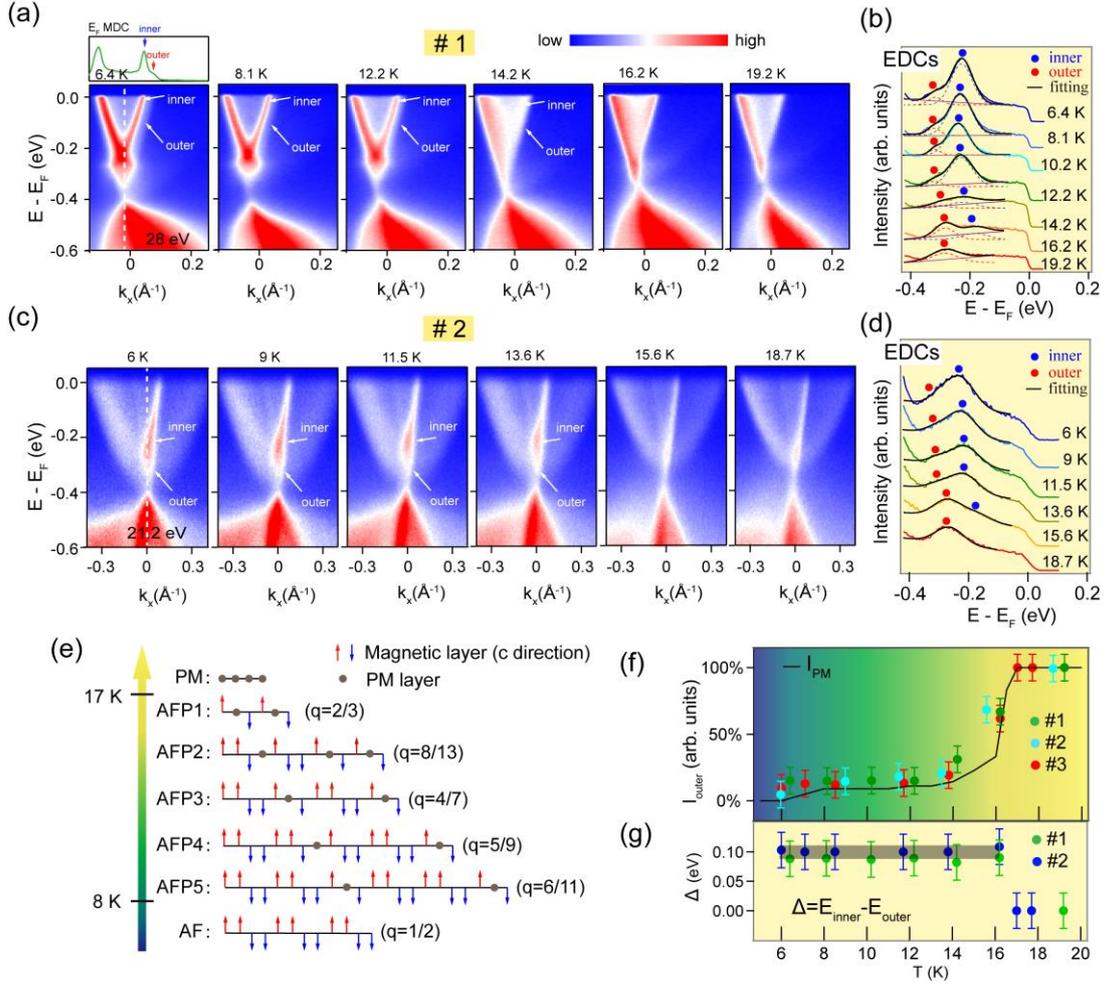

**Fig. 4 Temperature dependence of the electron bands in the *ab*-domain.** (**a, c**) Temperature-dependent ARPES spectra of the electron bands in the *ab*-domain, obtained using 28 eV and 21.2 eV photons, respectively. (**b, d**) Energy distribution curves (EDCs) extracted from **a** and **c** along the white dashed lines, respectively, with offset for clarity. Black lines represent fits to the EDC peaks. Red and blue dots indicate the peak positions of the outer and inner bands, respectively. Purple lines in b indicate the background distribution. (**e**) Schematic illustration of the devil's staircase. (**f**) Temperature dependence of the relative intensity of the outer bands ($I_{outer} = outer/(inner + outer)$) from three different samples. The calculated occupancy of the paramagnetic layers ($I_{PM}$), derived from the DFT+U calculations based on the devil's staircase model, is shown for comparison. (**g**) Temperature dependence of the energy separation between the inner and outer bands.